\documentclass[aps,pra,preprint,groupedaddress,showpacs,showkeys]{revtex4}
\usepackage{amssymb,bm}
\usepackage{amsmath}
\usepackage{graphicx}

\begin{document}
\bibliographystyle{apsrev}

\title{Kinetics of  polarization in non-relativistic  scattering}

\author{A.I. Milstein  }\email{A.I.Milstein@inp.nsk.su}
\author{S.G. Salnikov  }\email{salsergey@gmail.com}

 \affiliation{Budker Institute of Nuclear Physics, 630090 Novosibirsk, Russia}

\date{\today}

\begin{abstract}
An approach is developed, which essentially simplifies derivation of kinetic equation for  polarization in non-relativistic  scattering.
This approach is applicable for collision  of  projectile particles  with  a target for any spins of colliding particles.
The most detailed consideration is performed for  the case of spin $1/2$ projectile particle interacting with spin $1/2$ particle of a target.
The solution of the kinetic equation for the case of zero initial
polarization is presented.
\end{abstract}
\maketitle
\section{Introduction}
It is known that the collision of high energy beams of polarized particles gives additional very important information as compared with the case of unpolarized particles. However, to obtain high energy beam of heavy particles with noticeable polarization is hard problem \cite{PAX2005,PAX2009,PAX2009A}.
Besides,   theoretical description of the kinetics of polarization is also nontrivial even in non-relativistic case and needs its further development, see Review \cite{Artru2009}.
In Ref. \cite{MS2005}, the kinetics of the polarization buildup during the interaction of stored protons (antiprotons) with a polarized target was investigated. The kinetic equation was written in terms of the cross section corresponding to spin flip transition and spin non-flip transition.
In Ref. \cite{MS2005}, the quantization axes  was directed along the polarization vector $\bm\zeta_T$ of the target. If $\bm\zeta_T$ is parallel or perpendicular to the momentum of particles in the beam, and the initial polarization of these particles is zero, then it follows from the arguments of parity that the polarization, arising as a result of interaction, is directed along $\bm\zeta_T$.  Similar approach is used in Ref. \cite{MSS2008} at the consideration of polarization effects in non-relativistic electron- proton  scattering.
If  $\bm\zeta_T$ is not parallel or perpendicular to the momentum of particles in the beam, then it is necessary to take into account additional terms in the kinetic equation which  lead to rotation of polarization vector in the process of the polarization buildup \cite{Kolya2006}. The  kinetic equation for the density matrix was derived for the first time in Ref.~\cite{Pipkin1964} by solving the quantum Liouville equation and expressing its solution via a spin-dependent scattering amplitude. The equation obtained in  Ref.~\cite{Pipkin1964} was applied to the spin-exchange optical pumping,  and it was shown that the spin-exchange
collisions lead to  rotation of electron spin. The same  kinetic
equation was later re-derived in Ref.~\cite{baryshevsky1996}
and applied to the analysis of the spin evolution, see also \cite{baryshevsky2011,baryshevsky2012}.

In the present paper we develop an approach which essentially simplifies  derivation of kinetic equation describing spin evolution for any spin of the target and particles in the beam.
We consider the non-relativistic scattering  and  restrict ourselves for simplicity to the case of projectiles remaining   in the beam after scattering. This case corresponds to small-angle scattering of the projectile. For instance, our approach can be  directly applied to  interaction of heavy particle with the flow of polarized electrons. However, our approach can be easily generalized to more complicated cases.

\section{Kinetic equation}

Let us consider a particle with the spin $S_1$ interacting with a flow of polarized particles with the spin $S_2$. Below this flow is  referred to as a beam. We are going to investigate time  evolution of  $<\bm S_1>$. We assume for simplicity that the mass $M_1$ of the first particle is much larger than the mass $M_2$ of the second particle. In this case,  in the rest frame of the first particle it is possible to consider the first particle as a source of some potential depending on the spin operators  $\bm S_1$ and  $\bm S_2$. The operator $\mathcal O$ is some operator  constructed from the spin operators, so that $[\mathcal O, \bm r]=0$ and
$[\mathcal O, \bm p]=0$, where $\bm r$ is the relative coordinate vector and   $\bm p$ is the corresponding momentum. The wave function, which corresponds to the scattering problem, has the asymptotic  form at large distances,
\begin{equation}\label{WF}
\psi_{\bm k}(\bm r)=\left[e^{i\bm k\cdot\bm r}+\frac{e^{ikr}}{r}\,F\right]\chi_1\chi_2\,,
\end{equation}
where $\bm k$ is the initial momentum, $\chi_1$ and $\chi_2$ are the spin wave functions of the corresponding particles, the operator $F$ depends on  $\bm n_0=\bm k/k$,  $\bm n=\bm r/r$, and the spin operators. To describe kinetics of  polarization, we should use the wave packet normalized to unity. We introduce the wave function
\begin{equation}\label{WF1}
\Psi_{\bm k}(\bm r)=\frac{1}{\sqrt{V}}\,e^{-\lambda r}\,\psi_{\bm k}(\bm r)\,,
\end{equation}
where  $V$ is some normalization volume, and $\lambda$ is some real parameter which we tend to zero at the end of calculations. The factor $e^{-\lambda r}$ allows one to perform integration by parts in the matrix elements (to use hermiticity  of the operators), the system of units $\hbar=1$ is used. Then we have the usual equation
\begin{equation}\label{HE}
\frac{d}{dt}\int d\bm r\,\Psi_{\bm k}^+(\bm r){\mathcal O}_H\Psi_{\bm k}(\bm r)=
iN\int d\bm r\,\Psi_{\bm k}^+(\bm r)[H,{\mathcal O}_H]\Psi_{\bm k}(\bm r)\,,
\end{equation}
where ${\mathcal O}_H=e^{iHt}{\mathcal O}e^{-iHt}$ is the Heisenberg operator, $H$ is the Hamiltonian of the system,
$N=1/V$ is the density, and $[a,b]$ stands for the commutator of the operators $a$ and $b$. Using the relation $H\psi_{\bm k}(\bm r)=E\psi_{\bm k}(\bm r)$, we can write Eq. (\ref{HE}) as follows,
\begin{eqnarray}\label{HE1}
&&\frac{d}{dt}\int d\bm r\,\Psi_{\bm k}^+(\bm r){\mathcal O}_H\Psi_{\bm k}(\bm r)\nonumber\\
&&=iN\int d\bm r\,\psi_{\bm k}^+(\bm r)\left\{[e^{-\lambda r},H]{\mathcal O}_He^{-\lambda r}+
e^{-\lambda r}{\mathcal O}_H[e^{-\lambda r},H]\right\}\psi_{\bm k}(\bm r)\,.
\end{eqnarray}
The commutator $[e^{-\lambda r},H]$ is proportional to the small parameter  $\lambda$, and only the contribution of large distances $r\sim 1/\lambda$ in the integral over $\bm r$ can compensate this small parameter. Therefore, at  calculation of the commutator we can leave only the kinetic energy operator in the Hamiltonian,
\begin{equation}
[e^{-\lambda r},H]\approx [e^{-\lambda r},\frac{p^2}{2M}]=-\frac{i\lambda}{2M}\left(\bm p\cdot\bm n e^{-\lambda r}+
e^{-\lambda r}\bm n \cdot\bm p\right)\,,
\end{equation}
where $M\approx M_2$ is the reduced mass. We can also use the asymptotic form (\ref{WF}) of the wave function. Besides, in the term corresponding to  interference of the plane wave and the spherical wave, the main contribution to the matrix element is given by the small angle $\theta$ between vectors
$\bm n$ and $\bm k$, namely $\theta^2\sim 1/kr\sim \lambda/k$.  Finally  we obtain the kinetic equation
\begin{eqnarray}\label{HEfinal}
\frac{d}{dt}<{\mathcal O}>=vN\,\mbox{Sp}\left\{\rho(t)\left [\int d\Omega_{\bm n}F^+{\mathcal O}F-\frac{2\pi i}{k}
\Big(F^+(0){\mathcal O}-{\mathcal O}F(0)\Big)\right]\right\}\,.
\end{eqnarray}
Here $v=k/M$, $d\Omega_n$ is the differential of the solid angle corresponding to vector $\bm n$, $F(0)$ is the operator $F$ calculated at $\bm n=\bm n_0$, and $\rho(t)$ is the density matrix which describes the spin state of the system, trace is taken over spin indexes  of both particles. The density matrix equals to $\rho(t)=\rho_1(t)\rho_2$, where $\rho_1(t)$ is the  time dependent density matrix of  the first  particle and $\rho_2$ is the  time independent density matrix of the beam. The matrix $\rho_1(t)$
should be found as a result of solution of kinetic equation (see below). In (\ref{HEfinal}) we assume that $F(0)$ is finite quantity, otherwise it is necessary to introduce a regularization.

If we set ${\mathcal O}=1$ then we obtain the unitarity relation
\begin{eqnarray}\label{UR}
\mbox{Sp}\left\{\rho(t)\left [\int d\Omega_{\bm n}F^+F-\frac{2\pi i}{k}
\Big(F^+(0)-F(0)\Big)\right]\right\}=0\,.
\end{eqnarray}
Eqs.~(\ref{HEfinal}) and (\ref{UR}) are valid for arbitrary spins $S_1$ and $S_2$ of the particles.

\section{The case of a particle with $S_1=1/2$.}
Let $S_1=1/2$. Then it follows from Eq.~(\ref{HEfinal}) for ${\mathcal O}=\bm\sigma_1$ that
\begin{eqnarray}\label{zeta1}
&&\frac{d}{dt}\bm \zeta_1=vN\,\mbox{Sp}\left\{\rho(t)\left [\int d\Omega_{\bm n}F^+{\bm\sigma_1 }F-\frac{2\pi i}{k}
\Big(F^+(0){\bm\sigma_1 }-{\bm\sigma_1 }F(0)\Big)\right]\right\}\,,\nonumber\\
&&\rho_1(t)=\frac{1}{2}\left[1+\bm \zeta_1(t)\cdot\bm\sigma_1\right]\,,\quad \bm \zeta_1(t)=<\bm\sigma_1>\,,
\end{eqnarray}
where $\bm\sigma_1$  are the  Pauli matrices acting on the spin variables of the first particle. The unitarity relation (\ref{UR}) gives
\begin{eqnarray}\label{UR1}
&&\mbox{Sp}\left[\rho_2\int d\Omega_{\bm n}F^+F\right]=\mbox{Sp}\left[\frac{2\pi i}{k}\rho_2
\Big(F^+(0)-F(0)\Big)\right]\,,\nonumber\\
&&\mbox{Sp}\left[\rho_2\bm\sigma_1\int d\Omega_{\bm n}F^+F\right]=\mbox{Sp}\left[\frac{2\pi i}{k}\rho_2\bm\sigma_1
\Big(F^+(0)-F(0)\Big)\right]\,.
\end{eqnarray}

For a particle with the initial  polarization $\bm\zeta$  in a moment $t$, and the polarization $\bm\zeta_f$ measured by a detector,  the  cross section $\sigma$  has the form
\begin{eqnarray}\label{sigma}
&&\sigma=\mbox{Sp}\left[\rho(t)\int d\Omega_{\bm n}F^+\rho_fF\right]=\frac{1}{2}(A+\bm B\cdot\bm\zeta_f)\,,\nonumber\\
&&\rho_f=\frac{1}{2}\left[1+\bm \zeta_f\cdot\bm\sigma_1\right] \,,\nonumber\\
&&A=\mbox{Sp}\left[\rho(t)\int d\Omega_{\bm n}F^+F\right]\,,\quad
\bm B=\mbox{Sp}\left[\rho(t)\int d\Omega_{\bm n}F^+{\bm \sigma_1}F\right]\,.
\end{eqnarray}
As a result of scattering the polarization becomes  equal to $\bm\zeta'=\bm B/A$. Performing summation over $\bm\zeta_f$ and using the unitarity relation (\ref{UR}), we
write the total cross section $\sigma_{tot}$ as
\begin{eqnarray}\label{sigmatot}
&&\sigma_{tot}=A=A_0+\bm B_0\cdot\bm\zeta_1\,,\nonumber\\
&& A_0= \frac{\pi i}{k}\mbox{Sp}\left[\rho_2\Big(F^+(0)-F(0)\Big)\right]\,,\quad
\bm B_0= \frac{\pi i}{k}\mbox{Sp}\left[\rho_2\bm\sigma_1\Big(F^+(0)-F(0)\Big)\right]\,.
\end{eqnarray}
In terms of $\bm B$ and $\bm B_0$, Eq. (\ref{zeta1}) reads,
\begin{eqnarray}\label{zeta11}
&&\frac{d}{dt}\bm \zeta_1=vN\,\left\{\bm B-\bm B_0- \frac{\pi i}{k}  \mbox{Sp}\left[\rho_2(\bm\zeta_1\cdot\bm\sigma_1)
\Big(F^+(0){\bm\sigma_1 }-{\bm\sigma_1 }F(0)\Big)\right]\right\}\,,
\end{eqnarray}
It is convenient to write Eq.~(\ref{zeta1}) also in another form,
\begin{eqnarray}\label{zeta12}
&&\frac{d}{dt}\bm \zeta_1=vN\,\mbox{Sp}\Big\{\frac{1}{2}\rho(t)\int d\Omega_{\bm n}\Big([F^+,{\bm\sigma_1}]F+F^+[{\bm\sigma_1},F]\Big)\nonumber\\
&&+\frac{\pi}{k}\rho_2[\bm\zeta_1\times{\bm\sigma_1}]
\Big(F^+(0)+F(0)\Big)\Big\}\,.
\end{eqnarray}
The quantity $F$ can be written as
\begin{equation}
F=F_0+\bm\sigma_1\cdot\bm F_1\,,
\end{equation}
where $F_0$ and $F_1$ are operators acting on the spin variables of the second  particle. Then we obtain,
\begin{eqnarray}\label{zeta12final}
&&\frac{d}{dt}\bm \zeta_1=vN\,\mbox{Sp}_2\Big\{\rho_2\int d\Omega_{\bm n}\Big[\bm F_1^+(\bm F_1\cdot\bm\zeta_1)+(\bm F_1^+\cdot\bm\zeta_1)\bm F_1-2(\bm F_1^+\cdot\bm F_1)\bm\zeta_1\nonumber\\
&&+i[(F_0^+\bm F_1-\bm F_1^+F_0)\times\bm\zeta_1]-2i[\bm F_1^+\times\bm F_1]\Big]
-\frac{2\pi}{k}\rho_2[(\bm F_1^+(0)+\bm F_1(0))\times\bm\zeta_1]\Big\}\,,
\end{eqnarray}
where $\mbox{Sp}_2$ stands for trace over spin variables of particles from the beam (flow of polarized particles).

\subsection{The case  $S_2=0$.}
For  $S_2=0$ we have,
\begin{equation}
F=f_0+f_1\bm\sigma_1\cdot\bm\nu\,,\quad F_0=f_0\,,\quad \bm F_1=f_1\bm\nu\,,\quad \bm\nu=[\bm n\times\bm n_0]\,,
\end{equation}
where $f_0$ and $f_1$ are some functions of $x=\bm n\cdot\bm n_0$.
For $S_2=0$, the unitarity relations (\ref{UR1}) reduces to one nontrivial relation
\begin{eqnarray}\label{UR10}
\int d\Omega_{\bm n}[|f_0|^2+\nu^2|f_1|^2]=\frac{4\pi}{k}\mbox{Im}f_0(0)\,.
\end{eqnarray}

Using Eq. (\ref{zeta12final}) we arrive at the following equation describing spin relaxation,
\begin{eqnarray}\label{ppifinal}
&&\frac{d}{dt}\bm \zeta_1=-\omega [\bm \zeta_1+(\bm \zeta_1\cdot\bm n_0) \bm n_0]\,,\nonumber\\
&&\omega=vN\int d\Omega_{\bm n}\,\nu^2 |f_1|^2\,.
\end{eqnarray}
The solution of this equation reads
\begin{eqnarray}\label{ppisol}
&&\bm\zeta_1(t)=[\bm\zeta_1(0)-\bm n_0\,(\bm \zeta_1(0)\cdot\bm n_0)]e^{-\omega t}+
\bm n_0\,(\bm \zeta_1(0)\cdot\bm n_0)e^{-2\omega t}\,.
\end{eqnarray}
Thus, during relaxation  we have not only diminishing of $\zeta_1(t)$  but also rotation of the direction of $\bm \zeta_1(t)$.

\subsection{The case  $S_2=1/2$ .}
For  $S_2=1/2$,
\begin{eqnarray}\label{ep}
&& \rho_2=\frac{1}{2}\left[1+\bm \zeta_2\cdot\bm\sigma_2\right]\,,\quad
F=f_0+(f_1\bm\sigma_1+f_2\bm\sigma_2)\cdot\bm\nu+T^{ij}\sigma_1^i\sigma_2^j\,,\nonumber\\
&& F_0=f_0+f_2\bm\nu\cdot\bm\sigma_2\,,\quad F_1^i=f_1\nu^i+
T^{ij}\sigma_2^j\,.
\end{eqnarray}
Here  $\bm\sigma_2$ are the  Pauli matrices acting on the spin variables of the  particle from the  beam, $\bm \zeta_2$ is the time
 independent polarization of the beam. The functions, $f_0$, $f_1$, and $f_2$
depend on $x=\bm n_0\cdot\bm n$, and  the symmetric tensor $T^{ij}$ is constructed from the vectors $\bm n_0$ and $\bm n$.
Using these definitions, we obtain from Eq.~(\ref{zeta12final}),
\begin{eqnarray}\label{zeta12finalA}
&&\frac{d}{dt}\zeta_1^i=R^{ij}\zeta_1^j+[\bm\zeta_1\times\bm{\mathcal F}]^i+{\mathcal G}^i\,,\nonumber\\
&&R^{ij}=2vN\int d\Omega_{\bm n}\Big[|f_1|^2(\nu^i\nu^j-\nu^2\delta^{ij})+\mbox{Re}(T^{ia*}T^{ja})-T^{ab*}T^{ab}\delta^{ij}
\Big]\,,\nonumber\\
&&{\mathcal F}^i=vN\Big\{2\int d\Omega_{\bm n}\mbox{Im}\Big[f_0^*T^{ij}\zeta_2^j
+f_2^*f_1(\bm\nu\cdot\bm\zeta_2)\nu^i\Big]
+\frac{4\pi}{k}\mbox{Re}T^{ia}(0)\zeta_2^a\Big\}\,,\nonumber\\
&&{\mathcal G}^i = 2vN\int d\Omega_{\bm n}\epsilon^{ijk} \epsilon^{abc}T^{ja*}T^{kb}\zeta_2^c\,.
\end{eqnarray}
 Here we use the relations,
 $$ \int d\Omega_{\bm n}\,\chi(\bm n\cdot\bm n_0)T^{ij}\nu^k=0\,,\quad
\int d\Omega_{\bm n}  \epsilon^{abc}\mbox{Im}(T^{ia*}T^{jb})=0\,,\quad
\int d\Omega_{\bm n}\mbox{Im}(T^{ia*}T^{ib})=0\,,  $$
  valid for any function $\chi(x)$. These relations can be easily proved  using the representation of the tensor $T^{ij}$  \cite{Wolf1952},
\begin{equation}
T^{ij}=\delta^{ij}f_3+(n^in^j+n^i_0n^j_0)f_4+(n^in^j_0+n^i_0n^j)f_5\,,
\end{equation}
 where  $f_{3,4,5}$ are some functions of $\bm n\cdot\bm n_0$. For $S_2=1/2$,  the unitarity relations (\ref{UR1}) reduce  to two nontrivial relations
\begin{eqnarray}\label{UR11}
&&\int d\Omega_{\bm n}[|f_0|^2+\nu^2|f_1|^2+\nu^2|f_2|^2+T^{ab*}T^{ab}]=\frac{4\pi}{k}\mbox{Im}f_0(0)\,,\nonumber\\
&&\int d\Omega_{\bm n}\Big\{2\mbox{Re}(f_0^*T^{ab})+2\mbox{Re}(f_1^*f_2)\nu^a\nu^b-T^{ij*}T^{\alpha\beta}\epsilon^{i\alpha a}\epsilon^{j\beta b}\Big]
=\frac{4\pi}{k}\mbox{Im}T^{ab}(0)\,.
\end{eqnarray}

Then we obtain the  form of the tensor $R^{ij}$ and the vectors
 $\bm {\mathcal F}$ and $\bm {\mathcal G}$,
\begin{eqnarray}\label{RFG}
&&  R^{ij}= A_1\delta^{ij}+B_1n^i_0n^j_0\,,\nonumber\\
&&\bm {\mathcal F}=A_2\bm\zeta_2+B_2(\bm\zeta_2\cdot\bm n_0)\bm n_0\,,\nonumber\\
&&\bm {\mathcal G}=A_3\bm\zeta_2+B_3(\bm\zeta_2\cdot\bm n_0)\bm n_0\,,
\end{eqnarray}
where $A_i$ and  $B_i$ are some numbers. These numbers are expressed as
\begin{eqnarray}\label{AB}
&&A_1=-vN\int d\Omega_{\bm n}\Big[|f_1|^2\nu^2+T^{ia*}T^{ia}+n_0^in_0^jT^{ia*}T^{ja}\Big]\,,\nonumber\\
&&B_1=-vN\int d\Omega_{\bm n}\Big[|f_1|^2\nu^2+T^{ia*}T^{ia}-3n_0^in_0^jT^{ia*}T^{ja}\Big]\,,\nonumber\\
&&A_2=vN\Big\{\int d\Omega_{\bm n}\mbox{Im}\Big[f_0^*\Big(T^{ii}-n_0^in_0^jT^{ij}\Big)+f_2^*f_1\nu^2\Big]
+\frac{2\pi}{k}\mbox{Re}\Big[T^{ii}(0)-n_0^in_0^jT^{ij}(0)\Big]\Big\}\,,\nonumber\\
&&B_2=-vN\Big\{\int d\Omega_{\bm n}\mbox{Im}\Big[f_0^*\Big(T^{ii}-3n_0^in_0^jT^{ij}\Big)+f_2^*f_1\nu^2\Big]
+\frac{2\pi}{k}\mbox{Re}\Big[T^{ii}(0)-3n_0^in_0^jT^{ij}(0)\Big]\Big\}\,,\nonumber\\
&&A_3=-2vN\int d\Omega_{\bm n}\Big[n_0^in_0^jT^{ia*}T^{ja}-\mbox{Re}\Big(n_0^in_0^jT^{ij}T^{aa*}\Big)\Big]\,,\nonumber\\
&&B_3=2vN\int d\Omega_{\bm n}\Big[3n_0^in_0^jT^{ia*}T^{ja}-3\mbox{Re}\Big(n_0^in_0^jT^{ij}T^{aa*}\Big)
+T^{ii}T^{jj*}-T^{ij}T^{ij*}  \Big]\,.
\end{eqnarray}
Note that $A_1<0$ and $A_1+B_1<0$.

If $\bm\zeta_2=0$ then $\bm {\mathcal F}=\bm {\mathcal G}=0$, and  we obtain from Eq.~(\ref{zeta12finalA})
\begin{eqnarray}\label{ppisol1}
&&\bm\zeta_1(t)=[\bm\zeta_1(0)-\bm n_0\,(\bm \zeta_1(0)\cdot\bm n_0)]e^{A_1 t}+
\bm n_0\,(\bm \zeta_1(0)\cdot\bm n_0)e^{(A_1+B_1) t}\,,
\end{eqnarray}
so that we have depolarization. The component of $\bm\zeta_1$ parallel to $\bm n_0$ and the component of $\bm\zeta_1$  transverse to this vector diminish  with different rates.

Let $\bm\zeta_2\ne 0$ but $[\bm\zeta_2\times\bm n_0]=0$ ($\bm\zeta_2$ is parallel to $\bm n_0$). In this case it is easy to find that
\begin{eqnarray}\label{parallel}
&&\bm\zeta_1(t)=e^{A_1t}\left\{\cos(\omega t)\bm\zeta_1(0)+\sin(\omega t) [\bm\zeta_1(0)\times \bm n_0]\right\}\nonumber\\
&&+\left[e^{(A_1+B_1)t}-e^{A_1t}\cos(\omega t)\right]\,(\bm\zeta_1(0)\cdot\bm n_0)\,\bm n_0\nonumber\\
&&+\left[e^{(A_1+B_1)t}-1\right] \frac{A_3+B_3}{A_1+B_1}\,\bm\zeta_2\,,\quad \omega=(A_2+B_2)\zeta_2\,.
\end{eqnarray}

For $(\bm\zeta_2\cdot\bm n_0)=0$ ($\bm\zeta_2$ is perpendicular to $\bm n_0$), the solution reads
\begin{eqnarray}\label{perp}
&&\bm\zeta_1(t)=e^{(A_1+B_1/2)t}\cos(\Omega t)\bm\zeta_1(0)\nonumber\\
&& + \left\{\left[e^{A_1t}-e^{(A_1+B_1/2)t}\cos(\Omega t)\right]\frac{(\bm\zeta_1(0)\cdot\bm\zeta_2)}{\zeta_2^2}
 +\left[e^{A_1t}-1\right]\frac{A_3}{A_1}\right\}\,\bm\zeta_2  \nonumber\\
 &&+e^{(A_1+B_1/2)t}\frac{\sin(\Omega t)}{\Omega}\Bigg\{ A_2[\bm\zeta_1(0)\times\bm\zeta_2]\nonumber\\
&& +\frac{B_1}{2}\left[ (\bm\zeta_1(0)\cdot\bm n_0)\,\bm n_0-
\frac{(\bm\zeta_1(0)\cdot[\bm\zeta_2\times\bm n_0])}{\zeta_2^2}\,[\bm\zeta_2\times\bm n_0]\right]\Bigg\} \,,\nonumber\\
&&\Omega=\sqrt{A_2^2\zeta_2^2-B_1^2/4}\,.
\end{eqnarray}

For the general case, $[\bm\zeta_2\times\bm n_0]\ne 0$ and $(\bm\zeta_2\cdot\bm n_0)\ne 0$,  it is convenient to write $\bm\zeta_1(t)$ as
\begin{equation}
\bm\zeta_1(t)=(\alpha+\gamma)(\bm\zeta_2\cdot\bm n_0)\bm n_0-\beta(\bm\zeta_2\cdot\bm n_0)[\bm\zeta_2\times\bm n_0]-\gamma \bm\zeta_2\,,
\end{equation}
where $\alpha$, $\beta$, and $\gamma$ are some functions of time. For $\zeta_1^2$ we have
\begin{equation}
\bm\zeta_1^2(t)=\alpha^2(\bm\zeta_2\cdot\bm n_0)^2+\beta^2(\bm\zeta_2\cdot\bm n_0)^2[\bm\zeta_2\times\bm n_0]^2+\gamma^2 [\bm\zeta_2\times\bm n_0]^2\,.
\end{equation}
 From Eqs. (\ref{zeta12finalA}) and (\ref{RFG}) we find
\begin{eqnarray}
&&\frac{d\alpha}{dt}=(A_1+B_1)\alpha-A_2[\bm\zeta_2\times\bm n_0]^2\beta+A_3+B_3\,\nonumber\\
&&\frac{d\beta}{dt}=A_2\alpha+A_1\beta+(A_2+B_2)\gamma\,\nonumber\\
&&\frac{d\gamma}{dt}=-(A_2+B_2)(\bm\zeta_2\cdot\bm n_0)^2\beta+A_1\gamma-A_3\,.
\end{eqnarray}
Let us write these equations in the matrix form,
\begin{eqnarray}
&&\frac{d}{dt}\psi(t)=U\psi+\xi\,, \nonumber\\
&&\psi=\begin{pmatrix}
\alpha \\
\beta\\
\gamma
\end{pmatrix}\,,\quad
\xi=
\begin{pmatrix}
A_3+B_3\\
0\\
-A_3
\end{pmatrix}\,.
\end{eqnarray}
The formal solution of this equations has the form
\begin{equation}
\psi(t)=e^{Ut}\,\left[\psi(0)+U^{-1}\xi\right]-U^{-1}\xi\,.
\end{equation}
The eigenvalues $\Lambda_i$ of the matrix $U$ read $\Lambda_i=A_1-B_1\lambda_i$, where
\begin{equation}
\lambda_i^3+\lambda_i^2+(c_1+c_2)\lambda_i+c_2=0\,,\quad c_1=\frac{A_2^2}{B_1^2}[\bm\zeta_2\times\bm n_0]^2\,,\quad  c_2=\frac{(A_2+B_2)^2}{B_1^2}(\bm\zeta_2\cdot\bm n_0)^2\,.
\end{equation}
It is possible to show that all $\Lambda_i$ have negative real parts so that the asymptotic form of $\psi(t)$ at large $t$ is $\psi(t\rightarrow \infty)=-U^{-1}\xi$, or
\begin{eqnarray}
&&\alpha=-\frac{1}{W}\Large\{(A_2+B_2)[(A_2+B_2)(A_3+B_3)(\bm\zeta_2\cdot\bm n_0)^2+A_2A_3[\bm\zeta_2\times\bm n_0]^2]+A_1^2(A_3+B_3)\Large\}\,,\nonumber\\
&&\beta=\frac{1}{W}[A_1(A_2B_3-A_3B_2)-B_1A_3(A_2+B_2)]\,,\nonumber\\
&&\gamma=\frac{1}{W}[A_2(A_2+B_2)(A_3+B_3)(\bm\zeta_2\cdot\bm n_0)^2+A_3A_2^2[\bm\zeta_2\times\bm n_0]^2+A_1A_3(A_1+B_1)]\,,\nonumber\\
&&W=(A_1+B_1)(A_2+B_2)^2(\bm\zeta_2\cdot\bm n_0)^2+A_1A_2^2[\bm\zeta_2\times\bm n_0]^2+A_1^2(A_1+B_1)\,.
\end{eqnarray}
As should be, this asymptotic form is independent of the initial condition. It is also time independent though $\bm\zeta_1(t)$ changes its direction in a process of polarization at finite $t$.

\section{Conclusions}
We have  developed an approach which essentially simplifies  derivation of the kinetic equation describing spin evolution for any spin of the target and particles in the beam.
As an example, we have considered  the non-relativistic scattering of the flow of  polarized light particles on the heavy particle. For $S_1=S_2=1/2$, we have obtained the explicit solution of the  kinetic equation (\ref{zeta12finalA}), which is expressed via a few constants, Eq.~(\ref{AB}). The  asymptotic form of the solution is independent of the initial condition and time (there is no rotation at large time) though $\bm\zeta_1(t)$ changes its direction in a process of polarization. Our approach can be easily generalized to more complicated cases.

\section*{Acknowledgements}
The work  was supported by the Ministry of Education and Science of the
Russian Federation.

\end{document}